\documentclass[aps,prd,twocolumn,floatfix,showpacs]{revtex4}
\usepackage{graphicx}
\usepackage{amssymb}
\begin{document}

\title{{\bf Confining the Electroweak Model to a Brane}}

\author{Gonzalo A. Palma}

\email{G.A.Palma@damtp.cam.ac.uk}

\affiliation{\mbox{Department of Applied Mathematics and Theoretical
Physics, Centre for Mathematical Sciences, University of Cambridge,}
Wilberforce Road, Cambridge CB3 0WA, United Kingdom}

\date{February 2006}

%----------------------------------------------------------------------------Abstract
\begin{abstract}
We introduce a simple scenario where, by starting with a
five-dimensional $SU(3)$ gauge theory, we end up with several 4-D
parallel branes with localized fermions and gauge fields.  Similar
to the split fermion scenario, the confinement of fermions is
generated by the nontrivial topological solution of a $SU(3)$ scalar
field. The 4-D fermions are found to be chiral, and to have
interesting properties coming from their 5-D group representation
structure. The gauge fields, on the other hand, are localized by
loop corrections taking place at the branes produced by the
fermions. We show that these two confining mechanisms can be put
together to reproduce the basic structure of the electroweak model
for both leptons and quarks. A few important results are: Gauge and
Higgs fields are unified at the 5-D level; and new fields are
predicted: One left-handed neutrino with zero-hypercharge, and one
massive vector field coupling together the new neutrino with other
left-handed leptons. The hierarchy problem is also addressed.
\end{abstract}


\maketitle

%----------------------------------------------------------------------------Introduction

\section{Introduction} \label{S1}

One of the most remarkable twists that the braneworld scenario has
introduced in our view of physics is that the fundamental scale of
gravity could be significantly closer to scales currently accessible
by experiments than previously thought. In the braneworld paradigm,
the standard model of physics is localized to a four dimensional
brane while gravity (and possibly other fields) propagate in the
entire space, the bulk. In the 4-D perspective, this results in the
rescaling of many couplings and mass scales present in the theory,
thus providing an alternative approach to the hierarchy problem
\cite{Ant, Ant2, Ant3, Ant4, TeV1, TeV2}. Naturally, an important
problem in the study of this type of theories is understanding the
possible ways in which the standard model can be localized to a
brane \cite{Rev1, Rev2}; different mechanisms to localize matter and
gauge fields to a brane may have distinctive features with relevant
implications for braneworld phenomenology. In addition, several
aspects of the standard model's rich structure could be understood
in terms of how physics is arranged in the bulk.

A simple mechanism for the confinement of higher dimensional
fermions to a domain wall was proposed long ago by Rubakov and
Shaposhnikov \cite{fermions1} and is based purely on field
theoretical considerations. In their proposal, the wave functions of
fermion zero modes concentrate near the existing domain walls,
generating 4-D massless chiral fermions attached to them. This
mechanism has given rise to interesting braneworld scenarios with
clear consequences for physics beyond the standard model. One is the
split fermion scenario, proposed by Arkani-Hamed and Schmaltz
\cite{fermions2}. Here, bulk fermions are split into different
positions around the brane, offering a simple solution to the
hierarchy problem and the proton decay problem: the separation
between chiral fermions along the extra dimension generates
exponentially suppressed couplings between them (for example, Yukawa
couplings) \cite{fermions3, fermions4}.

In the case of gauge fields, a mechanism for their localization
(closely related to the confinement of fermions) is also available.
This is the case of the quasilocalization of gauge fields, proposed
by Dvali, Gabadadze and Shifman \cite{gauge} (see \cite{alt1, alt2}
for alternatives). Here, the interaction between bulk gauge fields
and the ``already'' localized fermions induces gauge kinetic terms
on the brane. The result is a 4-D effective theory consisting of
gauge fields mediating interactions between the localized fermions.
An interesting feature of this type of mechanism is the appearance
of a crossover scale $r_{c}$: at distances below this scale the
propagation of gauge fields along the brane is manifestly
four-dimensional, whereas above this scale the propagation becomes
five-dimensional.

In this paper we put together both types of confining mechanisms
---for fermions and gauge fields--- to reproduce the basic structure
of the electroweak sector of the standard model. We show that the
gauge symmetry exhibited by bulk fermions can be broken down through
their confinement to a domain wall, giving rise to non-trivial
subgroup representations. More precisely, by starting with a
five-dimensional $SU(3)$ gauge theory in the bulk, we obtain an
$SU(2) \times U(1)$ chiral theory on the brane, with all the basic
requirements of the electroweak model.

The key ingredient of the present proposal is that the positions at
which 5-D fermions end up localized depend on their $SU(3)$ charges.
This allows, for example, to break the $\mathbf{10}$ and $\bar
\mathbf{6}$ representations of $SU(3)$ down to the lepton and quark
representations of $SU(2) \times U(1)$, respectively, and confine
them to a single brane. In this construction it is possible to
identify the Higgs field with the fifth component of the localized
bulk gauge field. Additionally, new fields inevitably appear in the
resulting 4-D effective theory. These are: a left-handed neutrino
with zero-hypercharge, and a massive vector field coupling together
the new neutrino with other left-handed leptons.

This article is organized as follows: In Sec. \ref{S2} we introduce
the split fermion scenario and explain how the localization of
$SU(3)$ fermions to different positions in the bulk is produced.
Then, in Sec. \ref{S3} we analyze the confinement of gauge fields.
There we argue that the gauge symmetry of the localized fermions is
transferred to the gauge fields near the brane. Finally, in Sec.
\ref{S4}, we show that the electroweak model can be constructed by
putting these two mechanisms together. There, the hierarchy problem
is also addressed.

\section{Confinement of fermions} \label{S2}

%-----------------------------------------------------------------Confinement mechanism

In this section we describe the localization of bulk fermions to a
domain wall. We start with the split fermion scenario and then move
to a more complex setup where the localization of fermions depends
on their charges.

\subsection{Split fermions}

\begin{figure}[ht] %X-X-X-X-X-X-X-X-X-X-X-X-X-X-X-X-X-X-X-X-X-X-X-X-X-X-X-X- FIGURE 1
\begin{center}
\includegraphics[width=0.34\textwidth]{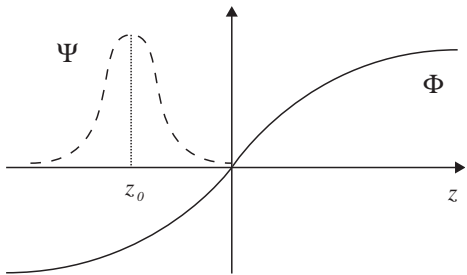}
\caption[Confinement]{The figure sketches the confinement of the
bulk fermion near the domain wall located at $z = 0$. The fermion
wave function is centered at position $z_{0} = - m \Delta^{2}$.}
\label{F1}
\end{center}
\end{figure}

Consider a 5-D system consisting of a spin-1/2 fermion $\Psi$ and a
real scalar field $\Phi$. To describe the 5-D space-time we use
coordinates $x^{A}$ with $A = 1, \ldots , 5$. The Lagrangian for the
system is
\begin{eqnarray}
\mathcal{L}^{(5)} = - \bar \Psi \left[\gamma^{A} \partial_{A}  + m +
y \, \Phi \right] \Psi  - \frac{1}{2} (\partial_{A} \Phi)^{2} -
V(\Phi). \nonumber\\ \label{eq2: L-split}
\end{eqnarray}
Here $m$ is the mass of the bulk fermion $\Psi$ and $y$ is a Yukawa
coupling. Additionally, $\gamma^{A}$ are the 5-D gamma-matrices in a
basis where
\begin{eqnarray}
\gamma^{5} = \left( \begin{array}{cc} 1 & 0 \\ 0 & -1 \end{array}
\right),
\end{eqnarray}
which is the usual four-dimensional $\gamma^{5}$ matrix. For the
time being we disregard the presence of gauge fields. Let us
consider the following potential for the scalar $\Phi$:
\begin{eqnarray}
V(\Phi) = \frac{\sigma}{4} \left[ \Phi^{2} - v^{2} \right]^{2}.
\end{eqnarray}
To discuss solutions to this system we use $z = x^{5}$ to
distinguish the extra-dimension and coordinates $x^{\mu}$ with $\mu
= 1, \cdots , 4$ to parameterize the usual 4-D space-time. Then, the
scalar field $\Phi$ is found to have a kink solution of the form:
\begin{eqnarray}
\Phi(z) = v \tanh \left( k z \right),
\end{eqnarray}
where $k = v \sqrt{\sigma / 2} \,$. The corresponding domain wall,
centered at $z=0$, is coupled to the fermion field through the
$y$-term. The equation of motion for $\Psi$ reads:
\begin{eqnarray}
\left[ \gamma^{\mu} \partial_{\mu} + \gamma^{5} \partial_{z} + m +
y \, \Phi (z) \right] \Psi = 0. \label{eq: Fermion-Wall}
\end{eqnarray}
Notice that the translational invariance along $z$ is broken. Thus,
in order to solve Eq. (\ref{eq: Fermion-Wall}) we define left and
right handed helicities $\Psi_{L}$ and $\Psi_{R}$, by $\gamma^{5}
\Psi_{L} = +\Psi_{L}$ and $\gamma^{5} \Psi_{R} = -\Psi_{R}$, and
expand them as:
\begin{eqnarray}
\Psi_{L,R} = \sum_{n} \Psi^{n}_{L,R} = \sum_{n} a^{L,R}_{n}(z) \,
\psi_{L,R}^{n}(x), \label{eq: expansion}
\end{eqnarray}
where $a_{n}^{L,R}(z)$ are Kaluza-Klein coefficients,
$\psi^{n}_{L,R}(x)$ are 4-D left and right-handed spinor fields, and
$n$ labels the expansion mode. Inserting the expansion (\ref{eq:
expansion}) back into Eq. (\ref{eq: Fermion-Wall}) we find the
following equations for the coefficients $a_{0} (z)$ and $a_{n} (z)$
with $n > 0$:
\begin{eqnarray}
\left[\pm \partial_{z} + m + y \, \Phi \right] a_{0}^{L,R} = 0,
\label{eq: zero} \\
\left[ -\partial_{z}^{2} + ( m + y \, \Phi)^{2} \mp y \,
(\partial_{z} \Phi) \right] a_{n}^{L,R} = \mu_{n}^{2} a_{n}^{L,R},
\label{eq: non-zero}
\end{eqnarray}
where $\pm$ stands for the left and right-handed helicities. At this
stage, it is convenient to define the following ``confinement''
length scale:
\begin{eqnarray}
\Delta = \frac{1}{\sqrt{|y v k|}}.
\end{eqnarray}
Then, in general, solutions to Eq. (\ref{eq: non-zero}) provide
modes with masses $\mu_{n}^{2}$ of order $\Delta^{-2}$. From now on
we assume that $\Delta$ is sufficiently small so that nonzero modes
can be integrated out without affecting the theory at low energies.
Solving Eq. (\ref{eq: zero}) the zero modes are found to be
\begin{eqnarray}
\Psi_{L,R} = A \exp \Big\{ \mp \! \int^{z}_{0} \!\!\! \left[ m + y\,
\Phi (z) \right] \, dz \Big\} \, \psi_{L,R}(x), \label{eq: solution}
\end{eqnarray}
where the factor $A$ is a normalization constant introduced in such
a way that
\begin{eqnarray}
\int  d z \, |\Psi|^{2} = |\psi(x)|^{2}.
\end{eqnarray}
Notice that only one of these two solutions is normalizable: if
$y>0$ ($y<0$) then the left (right) handed fermion is normalizable.
Additionally, observe that if $m=0$ then the fermion wave function
is centered at $z = 0$, otherwise its localization is shifted with
respect to the brane. To appreciate this, let us analyze the linear
behavior $\Phi \simeq v k z$ near $z = 0$ for the case $y>0$. Then,
if we assume that $m^{-1} \gg k \Delta^{2}$ (so the linear expansion
$\Phi \simeq v k z$ makes sense), we obtain
\begin{eqnarray}
\Psi_{L} \sim \frac{1}{\sqrt{\Delta}}  \exp \left[ - \frac{1}{2} \Delta^{-2} (z -
z_{0})^{2} \right] \, \psi_{L}(x),
\end{eqnarray}
where $z_{0} = - m \Delta^{2}$. Thus, the fermion wave function has
a width $\Delta$ and is centered at $z_{0}$. Figure \ref{F1}
sketches the confinement of the bulk fermion near the domain wall.
We can now compute the 4-D effective Lagrangian for $\psi_{L}(x)$ by
integrating out the extra-dimension:
\begin{eqnarray}
\mathcal{L}^{(4)} = - \bar \psi_{L} ( \gamma^{\mu}
\partial_{\mu} ) \psi_{L}. \label{eq2: L-split-eff}
\end{eqnarray}
Notice that in the limit $\Delta \rightarrow 0$ ($z_{0} \rightarrow
0$), we obtain a thin brane of the form:
\begin{eqnarray}
\mathcal{L}^{(5)} = \delta (z) \mathcal{L}^{(4)}.
\end{eqnarray}

There is an interesting consequence related to the shift of the
fermion's positions with respect to the domain wall: Suppose a
scenario in which two bulk fermions $\Psi^{1}$ and $\Psi^{2}$, with
masses $m_{1}$ and $m_{2}$, are coupled to a wall in such a way that
$y_{1} = y >0$ and $y_{2} = -y < 0$. If in the original 5-D
Lagrangian there is a term such as
\begin{eqnarray}
H \bar \Psi^{1} \Psi^{2} + \mathrm{h.c.},
\end{eqnarray}
where $H$ is a given bulk field (a scalar, for example), then the
4-D effective Lagrangian will contain a Yukawa term of the form:
\begin{eqnarray}
\sim  ( H \, \bar \psi^{1}_{L} \, \psi^{2}_{R} + \mathrm{h.c.} )
 \, e^{ -  r^{2} / 4 \Delta^{2} }, \label{eq: Yukawa-supp}
\end{eqnarray}
where $r = r_{1} - r_{2}$ is the separation between both fermion
wave functions with $r_{1} = - m_{1} \Delta^{2}$ and $r_{2} = +
m_{2} \Delta^{2}$. Physically, this means an exponential suppression
of the 4-D Yukawa coupling for the pair ($\psi^{1}_{L}$,
$\psi^{2}_{R}$) offering an interesting solution to the hierarchy
problem.

\subsection{Confining $SU(3)$ fermions}

We now proceed to analyze the localization of fermions produced by
``charged'' domain walls. Assume that space-time is described by a
5-D manifold $M$ with topology
\begin{eqnarray}
M = \mathbb{R}^{4} \times S^{1}, \label{eq: topology}
\end{eqnarray}
where $S^{1}$ is the one-dimensional circle and $\mathbb{R}^{4}$ is
the 4-D Lorentzian space. In this case, the coordinate $z = x^{5}
\in [0,L]$ is the spatial coordinate parameterizing $S^{1}$ with $L$
the size of the compact extra-dimension. Let us consider the
existence of 5-D bulk fermions transforming nontrivially under
$SU(3)$ gauge symmetry. They are described by the following
Lagrangian:
\begin{eqnarray}
\mathcal{L}_{\Psi}^{(5)} = - \bar \Psi [\gamma^{A} D_{A}  + Y(\Phi) ]
\Psi. \label{eq2: Lagrangian1}
\end{eqnarray}
The covariant derivative is $D_{A} \Psi = (\partial_{A} - i
E^{\alpha}_{A} T_{\alpha}) \Psi$, where $E^{\alpha}_{A}$ are $SU(3)$
bulk gauge fields. Here $\alpha = 1, \ldots , 8$ and $T_{\alpha}$
are the $SU(3)$ generators acting on $\Psi$. Observe that we are
considering a coupling term $Y(\Phi)$ where $\Phi = \Phi^{\alpha}
T_{\alpha}$ is a scalar field that transforms in the adjoint
representation of $SU(3)$. In order to construct
$SU(3)$-representations we proceed conventionally: We choose $T_{3}$
and $T_{8}$ as the Cartan generators and construct states to be
eigenvalues with charges:
\begin{eqnarray}
Q = (T_{3}, T_{8}).
\end{eqnarray}

Assume that $\Phi$ is dominated by the following $SU(3)$ gauge
invariant potential:
\begin{eqnarray}
V(\Phi) = \frac{\sigma}{4} \left[ \Phi^{\alpha} \Phi_{\alpha} -
v^{2} \right]^{2}.
\end{eqnarray}
Nonzero vacuum expectation solutions $\langle \Phi \rangle$ are
expected and, in general, they correspond to linear combinations of
$\langle \Phi^{3} \rangle$ and $\langle \Phi^{8} \rangle$.
Furthermore, since we are assuming the compact topology (\ref{eq:
topology}), then the system admits nontrivial topological solutions.
Take for instance the case of a single winding-number solution
\begin{eqnarray}
\langle \Phi (z) \rangle = \Phi_{0} \left[ \cos (k z) T_{3} +
\sin(k z) T_{8} \right], \label{eq: Phi(z)}
\end{eqnarray}
where $k = 2 \pi / L$ and $\Phi_{0}^{2} = v^{2} - k^{2}/\sigma$.
Notice that we have chosen $\langle \Phi^{8} \rangle = 0$ at $z =
0$. We can now proceed in the same way as before: we expand $\Psi$
in modes (\ref{eq: expansion}) and find zero mode solutions of the
form
\begin{eqnarray}
\Psi_{L,R} = A \exp \left\{ \mp \int^{z}_{0} \!\!\! Y (z) \, dz
\right\} \, \psi_{L,R}(x), \label{eq: solution2}
\end{eqnarray}
where $Y(z) \equiv Y[ \langle \Phi(z) \rangle ]$. To discuss the
consequences of solution (\ref{eq: Phi(z)}) with some transparency,
let us have a look to the following simple example: take a Yukawa
coupling of the form:
\begin{eqnarray}
Y(\Phi) = y \Phi =  y \Phi^{\alpha} T_{\alpha}, \label{eq: Yukawa}
\end{eqnarray}
and consider matter fields $\Psi$ belonging to the $\mathbf{3}$ [the
fundamental representation of $SU(3)$]. In this case the confinement
scale must be defined as
\begin{eqnarray}
\Delta = \frac{1}{\sqrt{|y \Phi_{0} k|}}.
\end{eqnarray}
Thus again, masses $\mu_{n}^{2}$ of nonzero modes solutions [Eq.
(\ref{eq: non-zero})] are found to be of order $\Delta^{-2}$.

\begin{figure}[ht] %X-X-X-X-X-X-X-X-X-X-X-X-X-X-X-X-X-X-X-X-X-X-X-X-X-X-X-X- FIGURE 2
\begin{center}
\includegraphics[width=0.3\textwidth]{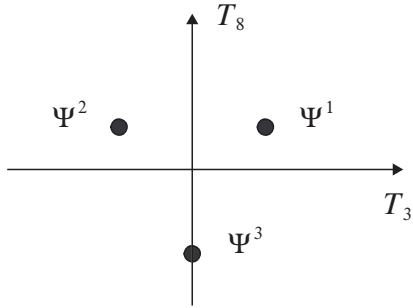}
\caption[Three]{The figure shows the $SU(3)$-charges, $T_{3}$ and
$T_{8}$, of fermions $\Psi^{i}$ (with $i = 1,2,3$) in the
fundamental representation $\mathbf{3}$.} \label{F2}
\end{center}
\end{figure}

\begin{figure}[ht] %X-X-X-X-X-X-X-X-X-X-X-X-X-X-X-X-X-X-X-X-X-X-X-X-X-X-X-X- FIGURE 3
\begin{center}
\includegraphics[width=0.4\textwidth]{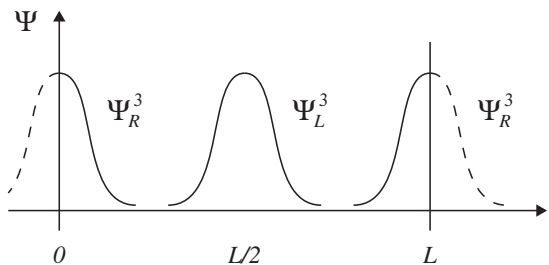}
\caption[Three conf]{The figure shows the way in which $\Psi^{3}$
is confined. Same representations but with different chiralities
end up in branes located at diametrically opposite positions in
the $S^{1}$ circle.} \label{F3}
\end{center}
\end{figure}

To work out the consequences of the Yukawa coupling (\ref{eq:
Yukawa}) on the $\mathbf{3}$ we chose $\Psi^{i}$ (with $i = 1,2,3$)
to have the following $SU(3)$-charges (see Fig. \ref{F2}):
\begin{eqnarray}
Q (\Psi^{1}) &=& (-1/2,+\sqrt{3}/6), \\
Q(\Psi^{2}) &=& (+1/2,+\sqrt{3}/6), \\
Q(\Psi^{3}) &=& (0,- \sqrt{3}/3).
\end{eqnarray}
In this way, replacing (\ref{eq: Yukawa}) into (\ref{eq:
solution2}), it is possible to see that the positions at which the
fermion wave functions end up centered depend on their
$SU(3)$-charges and their chirality. Observe, for instance, that in
the present realization left and right-handed fermions are localized
to diametrically opposite positions in the $S^{1}$ circle. Also, it
can be seen that if $|y \Phi_{0}| \gg k$, then the widths of the
fermion wave functions become of order $\Delta$ and the overlap
between fermions located at different positions becomes very small.
The following table provides the position of each state for the case
$y \Phi_{0} > 0$:
\begin{center}
\begin{tabular}{l c l | l c l}
Fermion  & \hspace{1em} & Position ($z$) \hspace{1em} & \hspace{1em} Fermion  & \hspace{1em} & Position ($z$) \\
\hline
\hspace{1em} $\Psi^{3}_{R}$ & & \hspace{1em} 0 &  \hspace{2em} $\Psi^{3}_{L}$ & & \hspace{1em} L/2 \\
\hspace{1em} $\Psi^{1}_{R}$ & & \hspace{1em} 2L/3 &  \hspace{2em} $\Psi^{1}_{L}$ & & \hspace{1em} L/6 \\
\hspace{1em} $\Psi^{2}_{R}$ & & \hspace{1em} 5L/6 &  \hspace{2em} $\Psi^{2}_{L}$ & & \hspace{1em} L/3 \\
\end{tabular}
\end{center}
Notice that the fundamental representation has been broken down to
several branes. Figure \ref{F3} shows the way in which $\Psi^{3}$ of
the fundamental representation is split.

We can now compute the 4-D effective theory for the matter fields
localized at any desired brane of our example. Let us compute, for
instance, the effective Lagrangian $\mathcal{L}_{\mathrm{eff}}$ at
the first brane ($z=0$) taking into account the presence of the
gauge field $E^{\alpha}_{A}$. In the limit $\Delta \rightarrow 0$
(with $L$ fixed), we obtain:
\begin{eqnarray}
\mathcal{L}_{\mathrm{eff}} = - \delta(z) \bar \psi^{3}_{R}
\gamma^{\mu} \bigg[
\partial_{\mu} + i \frac{\sqrt{3}}{3} E^{8}_{\mu} \bigg]
\psi^{3}_{R}. \label{eq: eff}
\end{eqnarray}
Here the delta function appears in the limit $\Delta \rightarrow 0$
after considering the right normalization factor $A$ in Eq.
(\ref{eq: solution2}). Notice the appearance of the induced current
\begin{eqnarray}
J_{8}^{\mu} = - i \frac{\sqrt{3}}{3} \bar \psi^{3}_{R}
\gamma^{\mu} \psi^{3}_{R},
\end{eqnarray}
which couples to the gauge field component $E^{8}_{\mu}$ in
(\ref{eq: eff}). The appearance of such currents will be important
to understand the localization of gauge fields (Sec. \ref{S2-Gen}).

\subsection{Generalization of the mechanism} \label{S2-Gen}

In general, given a nonzero v.e.v for a scalar field $\Phi (z)$, the
position $z$ at which the fermion wave function $\Psi$ is centered
is determined by the condition
\begin{eqnarray}
Y(z) \, \Psi = 0,
\end{eqnarray}
where $Y(z) = Y[\Phi (z)]$. The chirality of such a state is
determined by the sign of the derivative $\partial_{z} Y(\Phi)$ at
the given position. To be more precise, if $\partial_{z} Y(\Phi) >
0$ ($\partial_{z} Y(\Phi) < 0$), then the confined fermion is left
(right) handed.

\section{Localization of gauge fields} \label{S3}

%---------------------------------------------------------------------------Gauge sector

We now focus on the gauge sector of the model. The localization of
gauge fields to domain-walls is ensured by the already localized
fermionic fields; this is the case of the quasilocalization of gauge
fields \cite{gauge}. The interaction between the localized currents
at the branes with the 5-D gauge fields induces an effective 4-D
theory in the brane. This is produced by one-loop contributions to
the effective action coming from the brane currents.

\subsection{Quasi-localization of gauge fields}

For simplicity, we focus only on the localization of gauge fields to
the first brane ($z=0$) and neglect the effect of the coupling
between $E^{\alpha}_{A}$ and $\Phi$ on the 5-D behavior of
$E^{\alpha}_{A}$ near the brane. Now, assume that the spinor fields
are already confined and that the overlap between different branes
is very small ($\Delta^{-1} \gg k$). Then, in general, the
Lagrangian for the gauge fields $E^{\alpha}_{A}$ about the brane at
$z=0$ is given by
\begin{eqnarray}
\mathcal{L}^{(5)}_{\mathrm{G}} = -\frac{1}{4 g^{2}} F^{\alpha}_{A
B} F_{\alpha}^{A B} + \delta(z) E^{\alpha}_{A} J_{\alpha}^{A}(x)
\label{eq: gauge},
\end{eqnarray}
where $F^{\alpha}_{A B} = \partial_{A} E^{\alpha}_{B} -
\partial_{B} E^{\alpha}_{A} +  C^{\alpha}_{\beta \gamma}
E^{\beta}_{A} E^{\gamma}_{B}$ (here $C^{\alpha}_{\beta \gamma}$ are
the $SU(3)$ structure constants) and $g$ is the gauge coupling. As
mentioned, the currents $J^{A}_{\alpha}(x)$, localized at the
branes, appear as a consequence of the covariant derivative $D_A
\Psi = (\partial_A - i E_A^{\alpha} T_{\alpha}) \Psi$. To continue,
it is important to observe that, in general, the currents
$J^{A}_{\alpha}(x)$ do not continue transforming covariantly under
the full set of gauge symmetry transformations [as in Eq. (\ref{eq:
eff})]. This is because the many components of the $SU(3)$-spinor
representations end up at different positions along the fifth
dimension. In fact: since the effective terms for gauge fields are
induced by loops from these currents, then the transformation
properties of $J^{A}_{\alpha}(x)$ will be transferred to the
confined gauge fields. Take, for instance, the case of our previous
example in which the 4-D effective theory is given by Eq. (\ref{eq:
eff}). There, $\psi^{3}_{R}$ provides the current $J_{8}^{\mu} = - i
\frac{\sqrt{3}}{3} \bar \psi^{3}_{R} \gamma^{\mu} \psi^{3}_{R}$
which couples only to $E^{8}_{\mu}$. Then, a one-loop correction
induces the following Lagrangian for $E^{8}_{\mu}$ at the brane:
\begin{eqnarray}
\mathcal{L}^{(4)} = -\frac{1}{4 \lambda^{2}} (\partial_{\mu}
E^{8}_{\nu} - \partial_{\nu} E^{8}_{\mu})^{2},
\end{eqnarray}
where
\begin{eqnarray}
\lambda^{-2} = \frac{N}{12 \pi^{2}} \ln (\Lambda/\mu).
\end{eqnarray}
Here, $\Lambda$ and $\mu$ are the ultraviolet and infrared cut-offs
scales of the 5-D theory and $N = 1/3$ (which comes from the
coefficient $\sqrt{3}/3$ in $J_{8}^{\mu}$).

\subsection{Localization of $SU(2) \times U(1)$ gauge fields}

Let us now specialize to the case in which the localized currents
preserve the $SU(2) \times U(1)$ transformation properties at the first
brane $z=0$. Then it makes sense to perform the following
decomposition of the five-dimensional $SU(3)$ gauge field $E^{\alpha}_{A}$:
\begin{eqnarray}
W_{\mu}^{a} &=&  E_{\mu}^{a} \qquad  \mathrm{with} \qquad a =
1,2,3, \\
V_{\mu}^{i} &=&  E_{\mu}^{i} \qquad  \mathrm{with} \qquad i =
4,5,6,7,
\\
\phi^{i} &=&  E_{5}^{i} \qquad  \mathrm{with} \qquad i = 4,5,6,7,
\\
B_{\mu} &=&  E_{\mu}^{8}.
\end{eqnarray}
In the limit $\Delta \rightarrow 0$, other components of
$E_{A}^{\alpha}$ are decoupled from the matter fields confined to
the branes (this is because these components are coupling together
spinor fields with different chiralities that necessarily end up
at different branes). In this decomposition, the only non-zero
structure constant are: $C_{a b}^{c}$, $C_{i j}^{a}$ and $C_{i j
}^{8}$ (and obvious permutation of indices). Then, the current
term can be expressed as
\begin{eqnarray}
E^{\alpha}_{A} J^{A}_{\alpha} = W^{a}_{\mu} J_{a}^{\mu}(x) + B_{\mu}
J^{\mu}(x)  + V^{i}_{\mu} J_{i}^{\mu}(x) + \phi^{i} J_{i}(x), \nonumber\\
\end{eqnarray}
and the 4-D induced action for the now localized fields
$W_{\mu}^{a}$, $V_{\mu}^{i}$, $B_{\mu}$ and $\phi^{i}$ at the first
brane ($z=0$) becomes
\begin{eqnarray}
\mathcal{L}^{(4)}_{\mathrm{G}} &=& - \frac{1}{4 \lambda^{2}_{H}}
H^{a}_{\mu \nu} H_{a}^{\mu \nu} - \frac{1}{4 \lambda^{2}_{G}}
G_{\mu \nu} G^{\mu \nu} \nonumber\\ && - \frac{1}{2
\lambda^{2}_{\phi}} |D \phi|^{2}
 - \frac{1}{4 \lambda^{2}_{Q}} Q^{i}_{\mu \nu} Q_{i}^{\mu \nu}
+ \mathcal{L}_{V} . \qquad \label{eq: induced}
\end{eqnarray}
Here $H_{\mu \nu}^{a}$, $Q_{\mu \nu}^{i}$, $G_{\mu \nu}$ and
$D_{\mu} \phi^{i}$ are defined as
\begin{eqnarray}
H_{\mu \nu}^{a} &=&
\partial_{\mu} W_{\nu}^{a} -
\partial_{\nu} W_{\mu}^{a} +  C^{a}_{b c} W_{\mu}^{b}
W_{\nu}^{c}  ,  \nonumber\\
Q_{\mu \nu}^{i} &=& \partial_{\mu} V_{\nu}^{i} -
\partial_{\nu} V_{\mu}^{i} +  C^{i}_{a j} W_{\mu}^{a} V_{\nu}^{j}
+ C^{i}_{j a} V_{\mu}^{j} W_{\nu}^{a} \nonumber\\
&& + C^{i}_{8 j} B_{\mu} V_{\nu}^{j} + C^{i}_{j 8} V_{\mu}^{j}  B_{\nu} , \nonumber\\
G_{\mu \nu} &=& \partial_{\mu} B_{\nu} -
\partial_{\nu} B_{\mu}, \nonumber\\
D_{\mu} \phi^{i} &=& \partial_{\mu} \phi^{i} + C^{i}_{a j}
W^{a}_{\mu} \phi^{j} +  C^{i}_{8 j} B_{\mu} \phi^{j}  .
\end{eqnarray}
Additionally, in Eq. (\ref{eq: induced}) we have introduced
$\mathcal{L}_{V}$ which contains interaction terms between the
vector field $V_{\mu}^{i}$ and the rest of the induced fields
\begin{eqnarray}
\mathcal{L}_{V} =  - \frac{1}{4 \lambda^{2}_{1}} \left( R_{\mu \nu}^{a}
R^{\mu \nu}_{a} + K_{\mu \nu} K^{\mu
\nu}   \right)  - \frac{1}{2 \lambda^{2}_{2}} H^{a}_{\mu \nu}
R^{\mu \nu}_{a}  \nonumber\\  - \frac{1}{2 \lambda^{2}_{3}} G_{\mu \nu}
K^{\mu \nu} - \frac{1}{2 \lambda^{2}_{4}} \left( S_{\mu}^{a} S^{\mu}_{a}
+ S_{\mu} S^{\mu} \right), \qquad
\label{eq: L-V}
\end{eqnarray}
where we have defined: $R_{\mu \nu}^{a} =  C^{a}_{i j} V_{\mu}^{i}
V_{\nu}^{j}$, $S_{\mu}^{a} =  C^{a}_{i j} V_{\mu}^{i} \phi^{j}$,
$S_{\mu} = C_{i j}^{8} V^{i}_{\mu} \phi^{j} / \sqrt{3}$ and $K_{\mu
\nu} =  C^{8}_{i j} V_{\mu}^{i} V_{\nu}^{j} / \sqrt{3}$. Finally,
the various couplings $\lambda_{H}$, $\lambda_{G}$, $\lambda_{Q}$
and $\lambda_{\phi}$ in (\ref{eq: induced}), and $\lambda_{1}$,
$\lambda_{2}$, $\lambda_{3}$ and $\lambda_{4}$ in (\ref{eq: L-V})
are, in general, found to be of the form
\begin{eqnarray}
\frac{1}{\lambda^{2}} = \frac{N}{12 \pi^{2}} \ln
\frac{\Lambda}{\mu}, \label{eq: lambda}
\end{eqnarray}
where $N$ measures the number of fermions present in the different
loops, taking also into account the values of the various
$SU(3)$-charges and combinatorics. For example, we have
\begin{eqnarray}
N_{H} = \mathrm{Tr} \left( T_{3}^{2} \right), \quad \mathrm{and}
\quad N_{G} = \mathrm{Tr} \left( T_{8}^{2} \right),
\end{eqnarray}
where the traces run over all charged fermions taking place in the
loops inducing the first and second terms of (\ref{eq: induced}).
Notice, however, that the values of the $\lambda$-couplings may
change when taking into account the split of fermions. For instance,
as we shall see in Sec. \ref{hierarchy}, the $Y$ coupling of Eq.
(\ref{eq2: Lagrangian1}) could induce the split of fermions around a
single brane (for example, the first brane at $z=0$). This would
result in a modification of the way in which the induced 4-D
effective theory is computed, and therefore the way $N$ is computed
in (\ref{eq: lambda}). Nevertheless, the values of the
$\lambda$-couplings should all remain of the same order.

\subsection{Gauge theory near the brane}

The complete action describing the behavior of the gauge field
$E_{A}^{\alpha}$ near the first brane  now consists of:
\begin{eqnarray}
\mathcal{L}^{(5)}_{\mathrm{G}} = -\frac{1}{4 g^{2}} F^{\alpha}_{A
B} F_{\alpha}^{A B} + \delta(z) \mathcal{L}^{(4)}_{\mathrm{G}}
\label{eq: gauge2},
\end{eqnarray}
where $\mathcal{L}^{(4)}_{\mathrm{G}}$ is the induced Lagrangian
(\ref{eq: induced}). To study the propagation of gauge fields on
the braneworld it is convenient to define a crossover scale $r_{c}
= g^{2} / 2 \lambda^{2}$. Then, the physics taking place at the
brane can be shown to have two different regimes \cite{gauge}: at
large distances $r \gg r_{c}$ the propagator of the gauge fields
becomes five-dimensional, whereas at short distances $r \ll r_{c}$
it becomes four-dimensional.

%----------------------------------------------------------------------------Electroweak

\section{Confining the electroweak model to a brane} \label{S4}

We now turn to the confinement of the electroweak model. Our
approach consists of adding a new scalar field into the model so as
to allow a richer structure to the localization mechanism generated
by the $Y$-coupling. Then we show that leptons can be obtained from
the $\mathbf{10}$-representation of $SU(3)$, while quarks can be
obtained from the $\mathbf{\bar 6}$.

\subsection{Construction of the Electroweak brane}

To start, assume the existence of the same scalar field $\Phi =
\Phi^{\alpha} T_{\alpha}$ (as discussed previously) and an
additional scalar field $\Theta = \Theta^{\alpha} T_{\alpha}$ also
transforming in the adjoint representation of $SU(3)$. This scalar
is dominated by the following $SU(3)$ gauge invariant potential:
\begin{eqnarray}
U \propto \left[ \Theta^{\alpha} \Theta_{\alpha} - u^{2}
\right]^{2},
\end{eqnarray}
where $u$ is a constant parameter of the theory. Now, consider the
following $Y$-coupling:
\begin{eqnarray}
Y = - y  \left( \frac{1}{2} \{ \Phi , \Theta \} - \frac{1}{4}
\Theta^{\alpha} \Phi_{\alpha} + p \frac{\sqrt{3}}{2} u \, \Phi
\right) , \label{eq: Y}
\end{eqnarray}
where $\{ \, , \}$ denotes anticommutation. In the previous
equation, $p$ is a parameter of the model that depends on the
representation on which $Y$ is acting; in the present construction
we allow the value $p = 1$ if $Y$ couples to the $\mathbf{10}$, and
$p=-1/3$ if $Y$ couples to the $\mathbf{\bar{6}}$. Other gauge
invariant terms can also be included in (\ref{eq: Y}) without
modifying the main results of this section (we come back to this
point towards the end of this section).

We now focus on the case in which $\Theta$ acquires the following
v.e.v.:
\begin{eqnarray}
\langle \Theta \rangle = u \, T_{8}.
\end{eqnarray}
Then, after the scalars have acquired their respective v.e.v.'s we
are left with the following $z$-dependent coupling:
\begin{eqnarray}
(y \Phi_{0} u)^{-1} Y &=& - \left[ ( T_{8} + p \sqrt{3}/2 )T_{8} -
1/4 \right]
\sin(k z) \nonumber\\
&& -  \left[ T_{8} + p \sqrt{3}/2 \right] T_{3} \cos(kz). \label{eq: Y(z)gen}
\end{eqnarray}
Similar to our previous example, in this case the widths of the
fermion wave functions become of order $\Delta$ (the confining
length scale) which now is found to be
\begin{eqnarray}
\Delta = \frac{1}{\sqrt{|y \Phi_{0}| u k}}.
\end{eqnarray}
In what follows we analyze separately the confinement of leptons (from the
$\mathbf{10}$) and quarks (from the $\mathbf{\bar 6}$).

\subsection{Leptons}

Here we study the action of $Y$ on the $\mathbf{10}$ (where $p=1$)
and show that the confined fermions to the domain wall can be
identified with the usual leptons of the electroweak model.

\subsubsection{Confining leptons}

\begin{figure}[ht] %X-X-X-X-X-X-X-X-X-X-X-X-X-X-X-X-X-X-X-X-X-X-X-X-X-X-X-X- FIGURE 4
\begin{center}
\includegraphics[width=0.35\textwidth]{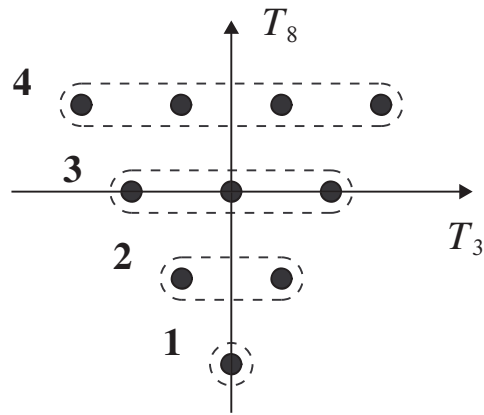}
\caption[Ten dec]{The figure shows the $\mathbf{10}$ representation
and its decomposition into $SU(2)$ subgroups: this is $\mathbf{10} =
\mathbf{1} \oplus \mathbf{2} \oplus \mathbf{3} \oplus \mathbf{4}$
with charges $T_{8} = -\sqrt{3}, -\sqrt{3}/2, \, 0, +\sqrt{3}/2$,
respectively.} \label{F4}
\end{center}
\end{figure}

\begin{figure}[ht] %X-X-X-X-X-X-X-X-X-X-X-X-X-X-X-X-X-X-X-X-X-X-X-X-X-X-X-X- FIGURE 5
\begin{center}
\includegraphics[width=0.35\textwidth]{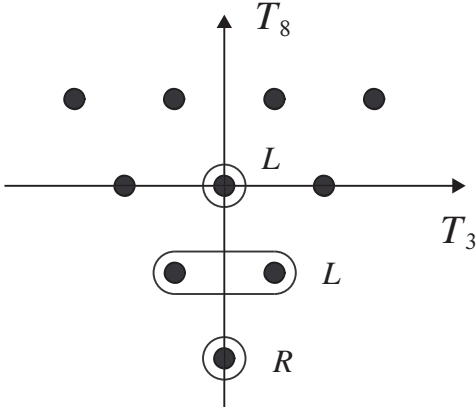}
\caption[Ten conf]{The figure shows those states of the
$\mathbf{10}$ that end up localized to $z=0$. The labels $L$ and
$R$ indicate the chirality of the confined states.} \label{F5}
\end{center}
\end{figure}

To proceed it is convenient to consider the decomposition of $SU(3)$
into $SU(2)$ subgroups (see Fig. \ref{F4}). The $\mathbf{10}$ has
the following decomposition: $\mathbf{10} = \mathbf{1} \oplus
\mathbf{2} \oplus \mathbf{3} \oplus \mathbf{4}$, with the following
$T_{8}$-charges: $T_{8} = -\sqrt{3}, -\sqrt{3}/2, \, 0,
+\sqrt{3}/2$.
Using this notation, we can work out the localization produced by
the $Y$-coupling to the first brane at $z=0$. First, observe from
Eq. (\ref{eq: Y(z)gen}) that all of those states in the
$\mathbf{10}$ with $(T_{8} + \sqrt{3}/2) T_{3} = 0$ give $Y=0$ at $z
= 0$. Then, following the reasoning of Sec. \ref{S2-Gen}, a chiral
fermion from each one of these states will confine to $z = 0$. The
precise chirality of each state depends on the sign of $\partial_{z}
Y(z)$. In the present case, assuming $y > 0$, the confined states
are: the right-handed $SU(2)$-singlet $R \equiv
\psi^{\mathbf{1}}_{R}$ with charge $Q=(0,-\sqrt{3})$; the two
left-handed components of the $SU(2)$-doublet $L \equiv
\psi^{\mathbf{2}}_{L}$ with charges $Q=(-1/2,-\sqrt{3}/2)$ and
$Q=(+1/2,-\sqrt{3}/2)$; and only one left-handed component from the
triplet $N \equiv \psi^{\mathbf{3}}_{L}$, with charge $Q=(0,0)$.
States with opposite chirality are confined to a ``mirror-brane''
located at $z=L/2$, and any other states are confined elsewhere.
Figure \ref{F5} shows those components of the $\mathbf{10}$ that
confine to $z=0$.

Now, the 4-D effective Lagrangian for the massless leptons at the
first brane is found to be
\begin{eqnarray}
\mathcal{L}_{\mathrm{lep}}^{(4)} &=& -  \bar L \Big[ \gamma^{\mu}
\partial_{\mu} - i  \gamma^{\mu} W^{a}_{\mu} T_{a}
+ i \frac{\sqrt{3}}{2} \gamma^{\mu} B_{\mu}    \Big] L \nonumber\\
&& - \bar R \Big[ \gamma^{\mu}
\partial_{\mu} + i \sqrt{3}  \gamma^{\mu} B_{\mu}   \Big] R
 -  \bar N \gamma^{\mu}
\partial_{\mu} N
 + \mathcal{L}_{\mathrm{I}}^{(4)}, \nonumber\\ && \label{eq: leptons}
\end{eqnarray}
where $\mathcal{L}_{\mathrm{I}}^{(4)}$ contains interaction terms
involving $\phi^{i}$ and $V^{i}_{\mu}$
\begin{eqnarray}
\mathcal{L}_{\mathrm{I}}^{(4)} =   - i \alpha \, \phi^{i} \bar R \, T_{i}
\, L - i \beta \, V^{i}_{\mu} \bar N \gamma^{\mu}
\, T_{i} \, L + \mathrm{h.c.},  \label{eq: leptons-I}
\end{eqnarray}
where $\alpha$ and $\beta$ are coefficients that appear from the
overlap between wave functions of different widths. In the present
case, $\alpha = \beta = (5)^{1/4} / \sqrt{3}$. In Eqs. (\ref{eq:
leptons}) and (\ref{eq: leptons-I}), $T_{a}$ and $T_{i}$ denote the
action of the corresponding $SU(3)$-generators on the
$SU(2)$-doublet $L = \psi^{\mathbf{2}}_{L}$. We can rewrite the
$T_{i}$'s in Eq. (\ref{eq: leptons-I}) to obtain a more transparent
notation
\begin{eqnarray}
\mathcal{L}_{\mathrm{I}}^{(4)} =   - i \alpha \frac{\sqrt{3}}{2}
\phi^{i} \bar R \, t_{i} \, L
 - i \beta V^{i}_{\mu} \bar N \gamma^{\mu}
\, s_{i} \, L + \mathrm{h.c.},  \label{eq: leptons-I2}
\end{eqnarray}
where $t_{i}$ and $s_{i}$ with $i = 4,5,6,7$, are $1 \times 2$
matrices acting on $L$ given by
\begin{eqnarray}
t_{4} = s_{6} = (1, 0),&& \qquad t_{5} = - s_{7} = i ( 1, 0),
\nonumber\\
t_{6} = s_{4} = (0, 1),&& \qquad t_{7} = - s_{5} = i (0, 1).
\end{eqnarray}

\subsubsection{Confining gauge fields}

The form of the theory presented in Eqs. (\ref{eq: leptons}) and
(\ref{eq: leptons-I}) corresponds to an $SU(2) \times U(1)$ gauge
theory with four massless chiral states. Therefore we can deduce the
quasilocalization of gauge fields to the first brane as discussed in
Sec. \ref{S3} [with the same Lagrangian shown in (\ref{eq:
induced})].

\subsubsection{Comparison with the electroweak model}

We can now compare this theory with the lepton sector of the
electroweak model. The two left-handed components $L$ and the
right-handed fermion $R$ can be identified with the usual
counterparts of the electroweak model, and $W^{a}_{\mu}$ and
$B_{\mu}$ with the $SU(2) \times U(1)$ gauge fields with couplings
$g_{1} = \lambda_{H}$ and $g_{2} = \sqrt{3} \lambda_{G}$
respectively. One of the most interesting aspects of this model,
however, is the appearance of two additional fields, namely the
vector field $V_{\mu}^{i}$ and the left-handed neutrino $N$ (which
has a zero-hypercharge). Observe that this neutrino interacts only
with the other left-handed particles $L$ through $V_{\mu}^{i}$.

If we further assume that $|\phi|$ develops a nonzero v.e.v.
$\phi_{0}$ (which can not be ruled out by symmetries), then
$\phi^{i}$ takes the role of the Higgs field. If this is the case,
two of the chiral states ($R$ and one of the $L$'s) mix together to
form an electron, while the other two remain massless (neutrinos).
The electroweak parameters are then found to be as follows: The
electron mass is $m_{e}^{2} = 3 \phi_{0}^{2} \lambda_{\phi}^{2} /
2$, the $W$-boson's mass is $M_{W}^{2} = \phi_{0}^{2}
\lambda_{H}^{2}/4$, and the electroweak angle is $\sin^{2}
\theta_{W} = 3 \lambda_{G}^{2}/(\lambda_{H}^{2} + 3
\lambda_{G}^{2})$. Very important for this model is that the
existence of $V^{i}_{\mu}$ has no conflicts with observations.
Fortunately, in the case of a nonzero v.e.v. $\phi_{0}$, the
four-component vector field $V_{\mu}^{i}$ becomes massive, with
$M_{V}^{2} = \phi_{0}^{2} \lambda_{\phi}^{2} \lambda_{Q}^{2} / 4
\lambda_{4}^{2}$.

%----------------------------------------------------------------------------Electroweak-Quarks

\subsection{Quarks}

\begin{figure}[ht] %X-X-X-X-X-X-X-X-X-X-X-X-X-X-X-X-X-X-X-X-X-X-X-X-X-X-X-X- FIGURE 6
\begin{center}
\includegraphics[width=0.35\textwidth]{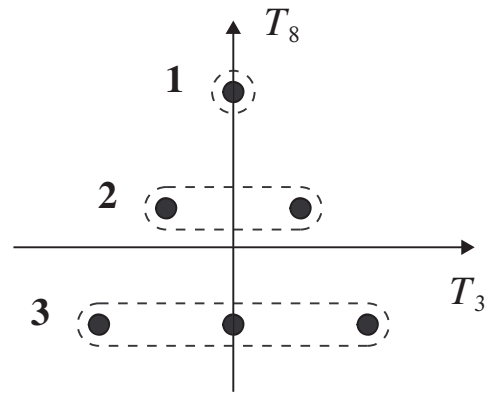}
\caption[Six dec]{The figure shows the $\mathbf{\bar 6}$
representation and its decomposition into $SU(2)$ subgroups: this is
$\mathbf{\bar 6} = \mathbf{1} \oplus \mathbf{2} \oplus \mathbf{3}$
with charges $T_{8} = +2 \sqrt{3}/3, + \sqrt{3}/6, -\sqrt{3}/3$,
respectively.} \label{F6}
\end{center}
\end{figure}

\begin{figure}[ht] %X-X-X-X-X-X-X-X-X-X-X-X-X-X-X-X-X-X-X-X-X-X-X-X-X-X-X-X- FIGURE 7
\begin{center}
\includegraphics[width=0.35\textwidth]{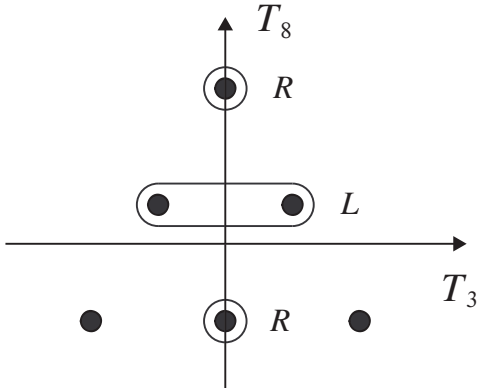}
\caption[Six conf]{The figure shows those states of the
$\mathbf{\bar 6}$ that end up localized to $z=0$. The labels $L$
and $R$ indicate the chirality of the confined states.} \label{F7}
\end{center}
\end{figure}

The case for quarks can be analyzed in exactly the same way as for
leptons. Here we need to consider the value $p = -1/3$ in the
$Y$-coupling. Having said this, recall that the $\mathbf{\bar 6}$
can be decomposed into $\mathbf{\bar 6} = \mathbf{1} \oplus
\mathbf{2} \oplus \mathbf{3}$ with the following $T_{8}$ charges:
$T_{8} = +2 \sqrt{3}/3, + \sqrt{3}/6, -\sqrt{3}/3$ (see Fig.
\ref{F6}).
Then, we obtain the following four massless chiral fermions confined
to the first brane: the right-handed $SU(2)$-singlet
$\psi_{R}^{\mathbf{1}}$ with charge $Q=(0,+2/\sqrt{3})$; the two
left-handed components of the $SU(2)$-doublet
$\psi_{L}^{\mathbf{2}}$ with charges $Q=(-1/2,+1/2\sqrt{3})$ and
$Q=(+1/2,+1/2\sqrt{3})$; and only one right-handed component from
the triplet $\psi_{L}^{\mathbf{3}}$, with charge
$Q=(0,-1/\sqrt{3})$. Figure \ref{F7} shows those components of the
$\mathbf{\bar{6}}$ that confine to $z=0$.
When the effective Lagrangian is computed we find the appropriate
quantum numbers for this sector to be identified with the quarks
of the standard model. A significant difference with the lepton
case, however, is the absence of interactions between quarks and
the vector field $V_{\mu}^{i}$.

\subsection{Solving the hierarchy problem} \label{hierarchy}

We have seen that the electron and $W$-boson masses are $m_{e} =
\sqrt{3/2} \phi_{0} \lambda_{\phi} $ and $M_{W} = \phi_{0}
\lambda_{H}/2$ respectively. What is more, the quark masses are
found to be proportional to $\phi_{0} \lambda_{\phi}$, of the same
order as the electron mass. This is just the hierarchy problem for
the particular case of the present model (recall that the
$\lambda$-couplings are all of the same order).

A simple way to correct this problem is to introduce a new term in
the definition of $Y$. For example, we could consider a new coupling
$Y'$ of the form:
\begin{eqnarray}
Y' = Y - y \, q \, v \, \Theta,
\end{eqnarray}
where $q$ is a dimensionless coefficient that could depend on the
representation on which $Y'$ is acting (observe the similarity of
the new term with the old one $- y \, p \, u \, \Phi$, in $Y$).
Then, after the scalars have acquired the v.e.v. discussed before,
the $Y'$ coupling becomes:
\begin{eqnarray}
Y'(z) = Y(z) - q (y v u ) T_{8}.
\end{eqnarray}
The second term of this expression resembles the 5-D mass term of
Eq. (\ref{eq2: L-split}). Therefore, the fermion wave functions will
be split around the branes and an exponential factor [like the one
of Eq. (\ref{eq: Yukawa-supp})] will appear suppressing the
couplings of Eq. (\ref{eq: leptons-I2}). This results in a hierarchy
between the mass scales of quarks, leptons and electroweak gauge
bosons.

Observe that in the definition of $Y$ we could also include terms
proportional to $\Phi^{2}$ and $\Theta^{2}$ with coefficients
depending on the representation. They would provide additional
terms contributing to the split of fermions around the brane.

\subsection{About the other branes}

\begin{figure}[ht] %X-X-X-X-X-X-X-X-X-X-X-X-X-X-X-X-X-X-X-X-X-X-X-X-X-X-X-X- FIGURE 8
\begin{center}
\includegraphics[width=0.37\textwidth]{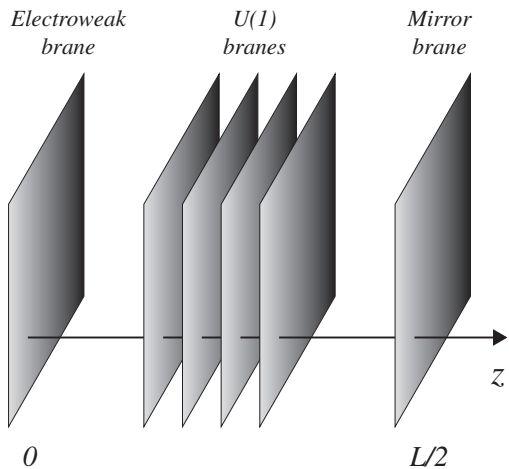}
\caption[Bulk configuration]{The figure sketches the disposition
of branes in the 5-D bulk. The electroweak brane is located at
$z=0$, while the mirror brane (a copy of the first brane but
containing matter with opposite chirality) is located at $z=L/2$.}
\label{F8}
\end{center}
\end{figure}

To finish, let us briefly mention that other branes are also formed
in the bulk. They appear from the localization of the rest of the
states in the $\mathbf{10}$ and $\mathbf{\bar 6}$ representations.
The most interesting brane is the ``mirror brane'' at $z=L/2$, which
contains a copy of the electroweak model obtained at the first brane
$z=0$ but with states having opposite chiralities. The rest of the
branes (also determined by the condition $Y=0$) all contain
different versions of $U(1)$ abelian gauge theories. Figure \ref{F8}
shows the 5-D configuration obtained in the construction.

%----------------------------------------------------------------------------Discussion

\section{Conclusions} \label{S5}

In this paper we have obtained a simple realization of the
electroweak model confined to a brane. The mechanism consisted in
breaking the $SU(3)$ gauge symmetry down in $SU(2) \times U(1)$
through the localization of bulk fermions to the brane. The
localization was produced by the coupling $Y$, of Eq. (\ref{eq: Y}),
between $SU(3)$ fermions and scalar fields with non-zero vacuum
expectation values. As in the split fermion scenario, the
four-dimensional fermions at the brane were found to be chiral. This
allowed us to achieve the electroweak chiral structure by localizing
those states (within given $SU(3)$-representations) with appropriate
charges to the same brane. For example, the lepton sector was
obtained from the $\mathbf{10}$ representation, while the quark
sector was obtained from the $\mathbf{\bar{6}}$.

Remarkably, in this model it was possible to identify
the Higgs field with the fifth component of the
$SU(3)$ bulk gauge field (see \cite{gh1, gh2, gh3, gh4, gh5, gh6}
for similar approaches). One problem with this result,
however, is the apparent difficulty in generating the appropriate
potential for the Higgs. Whether it is possible to obtain such a
potential in this particular setup remains an open question.

Another feature of the present construction is the presence of two
new fields coupled to the lepton sector of the standard model: A
four-component vector field $V_{\mu}^{i}$ (that transforms like the
Higgs under $SU(2) \times U(1)$ symmetry transformations) and a
left-handed neutrino $N$ (with zero-hypercharge). The existence of
these particles opens up interesting phenomenological possibilities.
For instance, the nonobservation of $V$-bosons pair-production at
LEP \cite{exp-W} is an indication of the constraint:
\begin{eqnarray}
 M_{V} > 104 \mathrm{GeV}.
\end{eqnarray}
Nevertheless, from the results of this paper we should not expect a
value $M_{V}$ significantly higher than $M_{Z}$ and $M_{W}$.  At the
same time, the mechanism generating the hierarchy between leptons,
quarks and gauge bosons, is also suppressing the couplings between
$V$ and leptons. If this is the case, then we could expect new
phenomena associated with extra-dimensions in lepton-collider
experiments in the near future.

Let us finish by mentioning that an important question that still
needs to be addressed within this model is how to include the
mixing between different families of leptons and quarks. For
instance, in the case of leptons, the new neutrino $N$ could be
playing some relevant role in the mixing of neutrinos.

%----------------------------------------------------------------------------Acknowledgments

\section*{Acknowledgements}

The author is grateful to Anne C. Davis, Daniel Cremades, David Tong
and Guy Moore for useful comments. This work is supported in part by
DAMTP (Cambridge) and MIDEPLAN (Chile).

%----------------------------------------------------------------------------References

\end{document}